\documentstyle[aps,preprint,floats,tighten]{revtex}

\input{psfig.tex}
\begin{document}
\draft

\preprint{{\sl Science} {\bf 280}, 1397 (1998)}

\title{The Case of the Curved Universe: Open, Closed, or Flat?}

\author{Marc Kamionkowski\footnote{Electronic address: \tt
kamion@phys.columbia.edu}}
\address{Department of Physics, Columbia University, 538 West
120th Street, New York, NY 10027}

\maketitle

\def\hatn{{\bf \hat n}}
\def\hatnprime{{\bf \hat n'}}
\def\hatnone{{\bf \hat n}_1}
\def\hatntwo{{\bf \hat n}_2}
\def\hatni{{\bf \hat n}_i}
\def\hatnj{{\bf \hat n}_j}
\def\vecx{{\bf x}}
\def\veck{{\bf k}}
\def\hatx{{\bf \hat x}}
\def\hatk{{\bf \hat k}}
\def\hatz{{\bf \hat z}}
\def\VEV#1{{\left\langle #1 \right\rangle}}
\def\cP{{\cal P}}
\def\noise{{\rm noise}}
\def\pix{{\rm pix}}
\def\map{{\rm map}}

\vskip 1cm

Determination of the geometry of the Universe has been a central
goal of cosmology ever since Hubble discovered its expansion
seventy-five years ago.  Is it a multidimensional equivalent of
the two-dimensional surface of a sheet of paper (``flat''), a
sphere (``closed''), or a saddle (``open'')?  The geometry
determines whether the Universe will expand forever or
eventually recollapse, and it may also shed light on its origin.
Particle theories suggest that in the extreme temperatures
prevalent in the very early Universe, gravity may have briefly
become a repulsive, rather than attractive, force.  If so, the
ensuing period of "inflation" \cite{inflation} could account for some
of the most fundamental features of the Universe, such as the
remarkable smoothness of the cosmic microwave background (CMB),
the afterglow of the big bang (see schematic timeline).

Until now, most astronomers have pursued the geometry by
attempting to measure the mass density of the Universe.
According to general relativity, if the density is equal to,
larger than, or smaller than a ``critical density'' fixed by the
expansion rate, then the Universe is flat, open, or closed,
respectively.  Several measurements currently seem to suggest a
density only a fraction $\Omega\simeq0.3$ of the critical
density (as opposed to $\Omega=1$ predicted by inflation).
However, most of these probe only the mass that clusters
with galaxies.  If a significant amount of some more diffuse
component of matter exists, such as neutrinos and/or ``vacuum
energy'' (Einstein's cosmological constant), then the
measurements do not necessarily tell us the geometry of the
Universe.  The research article by Gawiser and Silk
\cite{gawisersilk} on page 1405 of this issue and an
accompanying commentary on page 1398 by Primack tell this side
of the story \cite{primack}.

Another possibility is to look directly for the effects of
a curved Universe.  As an analogy, consider geometry on a
two-dimensional surface.  On a flat surface, the interior angles
of a triangle sum to 180 degrees and the circumference of a
circle is $2\pi$ times its radius.  However, when drawn on the
surface of a sphere, the interior angles of a triangle sum to
more than 180 degrees, and the circumference of a circle is less
than $2\pi$ times the radius.  Similar lines of reasoning show
that in an open (closed) Universe, objects of some fixed size
will appear to be smaller (larger) than they would in a flat
Universe.

\begin{figure}[htbp]
\centerline{\psfig{file=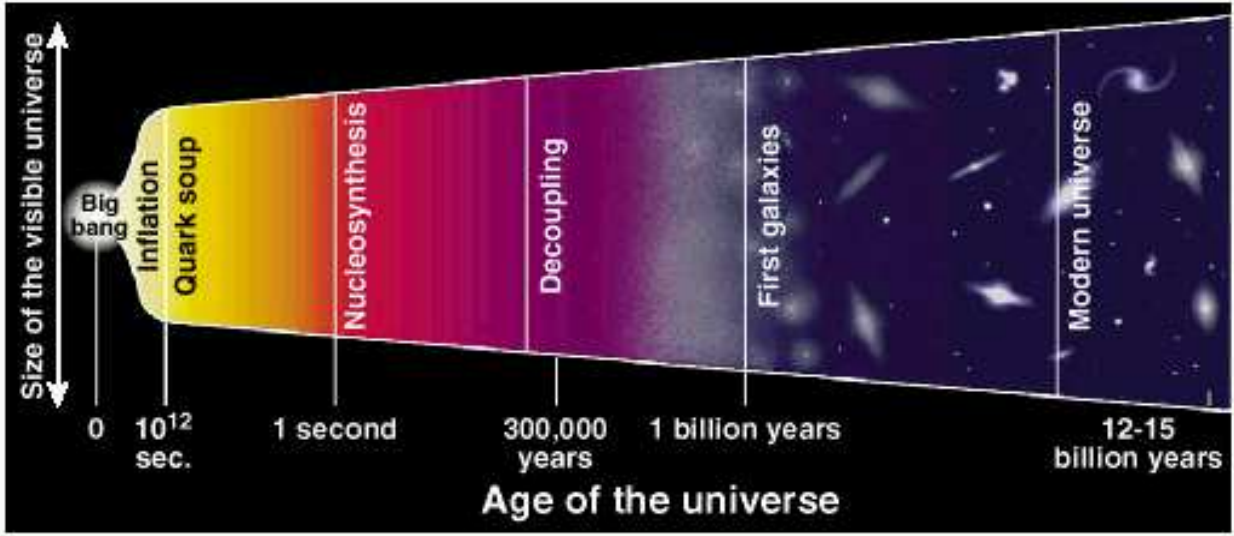,width=6.5in}}
\medskip
\caption{{\bf From smooth to structured.}  Schematic history of
     the Universe.  The big bang may have been followed by a period
     of rapid inflation, with the resulting ``soup'' of particles
     coalescing into nucleons and lighter elements.  Matter and
     radiation eventually became decoupled, the former
     gravitationally clumping into the structure of the modern
     Universe and the latter yielding the microwave background we see 
     today.  The seeds from which galaxies grew should be apparent in 
     the variations in the radiation background.
}
\label{figureone}
\end{figure}

The problem, then, is to find distant objects in the Universe of
known size (``standard rulers'').  It was recently
proposed that features at the CMB surface of last scatter could
provide such standard rulers \cite{kss}.  The photons that
make up the CMB last scattered roughly 10 to 15 billion years ago,
when the Universe was only 300,000 years old.  Therefore, when
we look at the CMB, we see a spherical surface in the early
Universe 10 to 15 billion light-years away.  Although galaxies and
clusters of galaxies had not yet formed, the seeds which later
grew into these structures existed, and we know the distribution
of their intrinsic sizes.  By measuring the distribution of
their apparent sizes on the sky, we can determine the
geometry of the Universe.

More precisely, one must measure
the angular power spectrum of the CMB:  Suppose we
measure the temperature $T(\vec\theta)$ as a function of direction
$\vec\theta$ on the sky over some approximately square region of
the sky.  We may then compute the Fourier transform
$\tilde(\vec\ell)$ of this temperature map.
The power spectrum is then
given by the set of multipole moments $C_\ell = \VEV{\tilde T(\vec \ell)
\tilde T^*(\vec\ell)}$, where the angle brackets denote an average
over all wavevectors $\vec\ell$ of magnitude $|\vec\ell|=\ell$.
Roughly speaking, each $C_\ell$ measures the mean-square
temperature difference between two points separated by an angle
$(\theta/{\rm deg})\simeq(200/\ell)$, so larger-$\ell$ modes
measure temperature fluctuations on smaller angular scales.
Increasingly accurate measurements of the $C_\ell$'s requires
mapping larger portions of the sky to reduce the sampling
error.  Precise temperature measurements are also required.
Good angular resolution is needed to determine the larger-$\ell$
moments.

\begin{figure}[htbp]
\centerline{\psfig{file=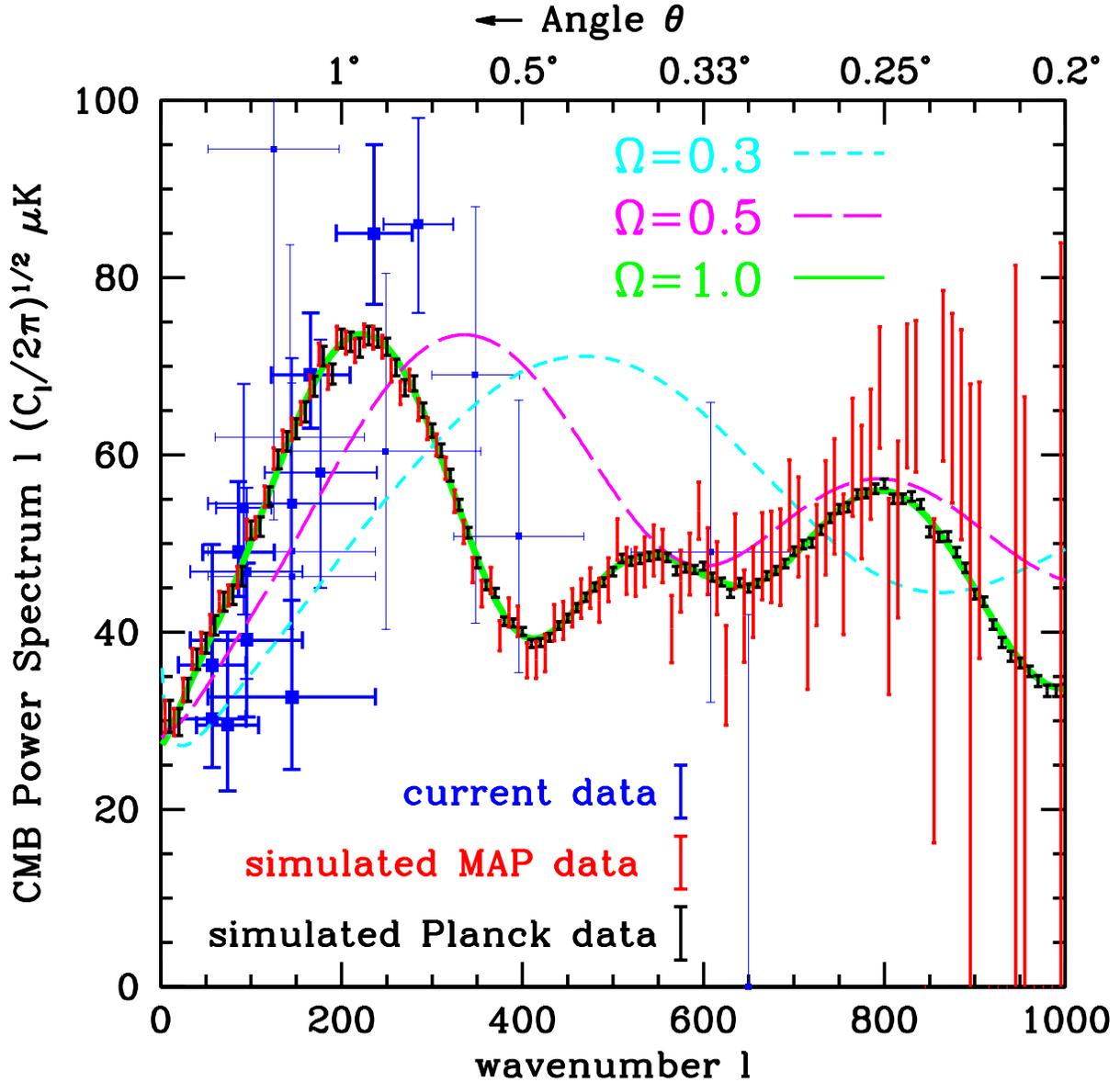,width=6.5in}}
\caption{{\bf Bumps in the background.}  Power spectrum of the
cosmic microwave background as a function of angle $\theta$ or
wavenumber $\ell$.  Curves show spectral behavior expected for
different mass densities, $\Omega$.  Future MAP data (simulated, 
red) should permit better constraints on which curve actually
represents the cosmic microwave background.  Even better
constraints should be produced by the future Planck Surveyor
mission (simulated, black).
}
\label{figuretwo}
\end{figure}

If galaxies and clusters grew from gravitational instability of
tiny primordial density perturbations, then the CMB power
spectrum (the $C_\ell$) should look like the curves shown in the 
graph. The bumps in the curves are due to physical
processes that lead to large-scale structures.  If $\Omega$ is
smaller than unity, then the Universe is open and the structure
in the CMB is shifted to smaller angular scales, or
equivalently, larger $\ell$'s. Therefore, the location of the
peaks (primarily the first peak) in the CMB spectrum determines
$\Omega$ and therefore the geometry of the Universe \cite{kss}.

The blue points are current measurements from balloon-borne and
ground-based experiments.  Several groups \cite{groups} have
recently found a value of $\Omega$ consistent with unity by
fitting these data to the theoretical curves.  Although these
results are intriguing and perhaps suggestive, even a cursory
glance demonstrates that the current data cannot robustly
support a flat Universe.

However, a new generation of experiments will soon provide
significant advances.  As indicated by the red points in the
Figure, the Microwave Anisotropy Probe (MAP), a NASA satellite
mission scheduled for launch in the year 2000, should confirm the
peak structure suggested by the gravitational-instability
paradigm (if it is correct) and make a precise determination of
the geometry.  The Planck Surveyor, a European Space Agency
mission scheduled for launch in 2005, should improve on MAP's
precision and may also illuminate the nature of the missing
mass.

If the peak structure of gravitational instability is confirmed
and the measurements are precisely consistent with the
inflationary prediction of a flat Universe, then new
avenues of inquiry will be opened to provide clues to the new
particle physics responsible for inflation.  As one example, the
polarization of the CMB may probe a stochastic background of
gravitational waves predicted by inflation \cite{polarization}.

\bigskip

This work was supported by D.O.E. contract DEFG02-92-ER 40699,
NASA contract NAG5-3091, and the Alfred P. Sloan Foundation

\end{document}